\documentclass[onecolumn,authoryear]{els-mrw} 

\usepackage{amsmath,amssymb,amsfonts,amsthm,makeidx,graphicx}
\usepackage{txfonts}
\usepackage{helvet}
\usepackage{aas_macros}

\usepackage{hyperref}
\usepackage{xcolor}

\definecolor{Blue}{rgb}{0,0.08,0.65}
\definecolor{Blue2}{rgb}{0,0.4,0.6}

\hypersetup{
    colorlinks=true,
}

\begin{document}

\chapter{Large Scale Structure and the Cosmic Web}\label{chap1}

\author[1]{Rita Tojeiro}%
\author[2]{Katarina Kraljic}%

\address[1]{\orgname{University of St Andrews}, \orgdiv{School of Physics and Astronomy}, \orgaddress{North Haugh, St Andrews, KY16 9SS, United Kingdom}}
\address[2]{\orgname{Observatoire Astronomique de Strasbourg}, \orgdiv{Universit\'e de Strasbourg, CNRS, UMR 7550}, \orgaddress{F-67000 Strasbourg, France}}

\articletag{Chapter Article tagline: update of previous edition,, reprint..}

\maketitle




\begin{glossary}[Nomenclature]
\begin{tabular}{@{}lp{34pc}@{}}
AGN & Active galactic nuclei \\
HOD & Halo occupation distribution \\
$\Lambda$CDM & $\Lambda$ cold dark matter \\
LSS & Large-scale structure \\
(s)SFR & (specific) Star formation rate \\
SHMR & Stellar-to-halo mass relation \\
TTT & Tidal torque theory \\
WHIM & Warm-hot intergalactic medium \\
\end{tabular}
\end{glossary}

\begin{abstract}[Abstract]
The formation and evolution of galaxies cannot be separated from large scale structure growth. Dark matter halos (and, therefore, galaxies) form and grow within the cosmic web – the classification of large-scale structure as distinct environments, namely voids, walls, filaments and nodes. Thanks to the rapid development of extragalactic spectroscopic redshift surveys and cosmological simulations over the last two decades, we are now able to measure the impact of the cosmic web on galaxies and halos in observations and in simulations. In this chapter we summarise the state of play in our understanding of the link between dark matter halos, galaxies, and the cosmic web.
\end{abstract}

\textbf{Key points}\\

\begin{itemize}

\item On large scales matter is organised into nodes, filaments, walls and voids, forming the so-called cosmic web.
\item The cosmic web originates from the initial fluctuations in the primordial density field, enhanced by anisotropic gravitational collapse at later times.
\item Simulations show that the cosmic web impacts on the properties of dark matter halos, at fixed halo mass: halos near or within large filaments are older (they have a lower accretion rate), have a lower substructure fraction, and have preferentially tangential orbits when compared to halos far from large filaments.
\item The growth of galaxies is linked to the growth of dark matter halos. Although this relationship is broadly understood, there remains uncertainty on the impact of halo growth on the age, colour, or star-formation rate of galaxies at a given halo mass. This makes it difficult to predict from first principles what the impact of the cosmic web on galaxy properties should be.
\item In the local Universe, observations are able to isolate the impact of the cosmic web on galaxies: e.g. galaxies that are closer to filaments are more massive, have lower sSFR and higher (stellar and gas-phase) metallicity. 
\item The impact of the cosmic web on HI fraction is debated in observation, potentially due to selection biases. 
\item The orientation of angular momentum and shape of observed galaxies depend on their position within the cosmic web. Massive and/or early-type galaxies tend to have their angular momentum perpendicular to the direction of the neighbouring filament, while for low mass and/or late-type galaxies, their spin is found to be aligned with their closest filament. Simulations are in qualitative agreement with these results.
\item The multi-scale aspect of the cosmic web is important, with different sized filaments or nodes having potentially very different roles in the evolution of galaxies and dark matter halos. 
\item Hydrodynamic simulations provide overall support for the observed impact of the cosmic web on galaxies, however, more detailed analysis is needed to understand smaller remaining discrepancies.
\end{itemize}

\section{Introduction}

Within the $\Lambda$ Cold Dark Matter ($\Lambda$CDM) framework, the formation and evolution of galaxies is intrinsically linked to structure formation and the matter distribution in the Universe. As structure formation is a process driven almost entirely by gravity, it is now well understood and accurately modelled using N-body simulations. The strongest link between galaxies and structure formation can be understood in terms of dark matter halos – virialised dark matter overdensities that decouple from the Hubble expansion and undergo gravitational collapse. Dark matter halos then provide the gravitational potential wells where cool gas is able to gather and form stars. This connection between galaxies and halos is at the heart of all models of galaxy formation in $\Lambda$CDM – it emerges from cosmological hydrodynamical simulations and is explicitly modelled in semi-analytic and empirical models. 

\begin{figure}
    \centering
    \includegraphics[width=1\linewidth]{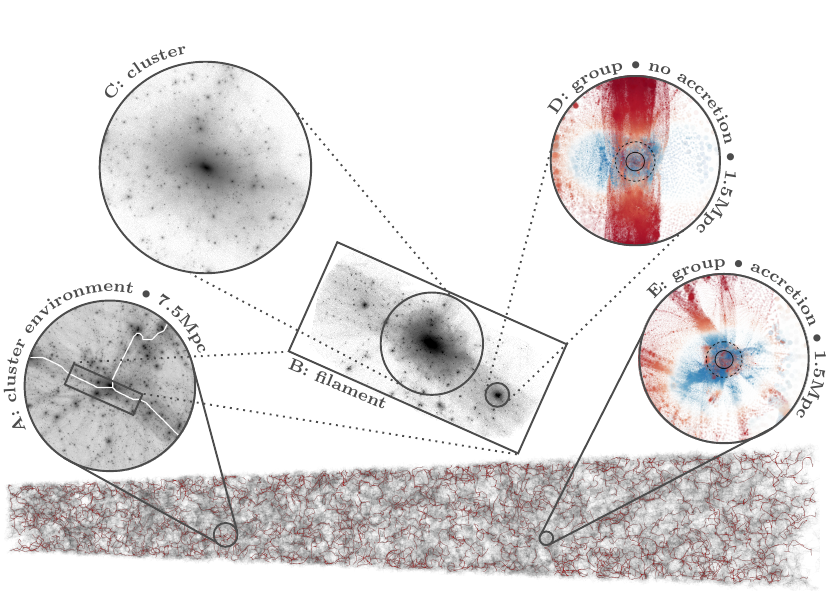}
    \caption{The cosmic web is the multi-scale classification of the complex and anisotropic distribution of matter on scales larger than a halo into distinct environments: voids, walls, filaments, and nodes. The light cone shows the matter distribution (grey) in the Euclid Flagship simulation, with the cosmic filaments as traced by the distribution of galaxies shown in red. Panel A shows the anisotropic environment in the outskirts a typical cluster from a higher resolution simulation. The anisotropic nature of the environment is lost if we focus on the central regions of the cluster alone (panel C). Panel B shows an intermediate view, where a cluster and groups are put in context of massive filaments, which can have a strong impact on the growth and evolution of halos. This impact is shown in panels D and E, showing two halos in different cosmic web environments, directly resulting in two different accretion and dynamical states. The colour scheme in panels D and E indicates whether material is infalling (blue) or receding (red) from the halo. Euclid Flagship simulation \citep{Castander-2024}, 
    courtesy of the Euclid Consortium.
    Neither panel is a direct zoom-in from the light cone, and are taken as typical examples only. Panels A, B, and C courtesy of Kyungwon Chun and show the same region at different levels of zoom from a higher-resolution simulation (the \protect{\href{https://darwin-simulation.github.io/}{Darwin}} project). Panels D and E adapted from \cite{Borzyszkowski-2017}.}
    \label{fig:figure-1}
\end{figure}

Whereas it is certainly possible - and useful - to understand the broad strokes of galaxy evolution within the context of dark matter halos alone, it has become increasingly clear that the full matter density field brings valuable additional information about the properties and evolution of halos and galaxies. The cosmic web – the classification of the matter density field in terms of distinct environments (voids, walls, filaments, nodes) – captures specific aspects of the matter density field that dark matter halos alone do not (Fig. \ref{fig:figure-1}) and has attracted substantial interest over the last decade or so in the context of galaxy evolution. Filaments have long been recognised as important structures that act as conduits of matter transfer. On the galaxy scale, thin filaments are associated with cold gas that feeds galaxies from the larger environment. On larger scales, filaments transfer matter towards nodes, where they intersect with other filaments. Over 40\% of dark matter is thought to be in large filaments at present day, and shock-heated diffuse gas in these large filaments is an equally important component of the baryon content of the Universe. It is important to understand the critical multi-scale aspect of filaments and the cosmic web in general – massive galaxy clusters are connected to and fed by large filaments, but the process happens also on a group and galaxy scale (see panels A and E of Fig.~\ref{fig:figure-1}). The other distinct feature of the cosmic web is its anisotropic nature, and many of the cosmic web features are simply lost if we consider large-scale structure, or environment, on spherically averaged terms only (see panels B and C of Fig.~\ref{fig:figure-1}). The focus here, therefore, is on trying to understand the additional physical processes that galaxies and dark matter halos undergo due to the anisotropic matter distribution and velocity flows within their vicinity. 

In this chapter we focus on the large-scale filaments and nodes as traced by galaxies, and their impact on halos and galaxies. Given the role of dark matter halos in understanding structure formation and its link with galaxy evolution, we begin by summarising the formation and growth of dark matter halos and its impact on galaxies (Section~\ref{sec:halos-to-galaxies}). We will see that isolating the impact of halo growth on the properties of galaxies remains an open question, both in simulations and observations. In Section~\ref{sec:cosmic-web} we introduce the cosmic web. We then consider the impact of the cosmic web specifically on the properties of dark matter halos and galaxies in Section~\ref{sec:CW-halos-galaxies}. We will see that dark matter halos are strongly impacted by their cosmic web environment, particularly in terms of their assembly history (Section~\ref{sec:CW-halos}) and spin (Section~\ref{sec:CW-galaxies-spin}). The impact on galaxies' stellar mass, colour, stellar content, and gas content is discussed in Section~\ref{sec:CW-galaxies-scalar}. The focus of this chapter is on large scale filaments and nodes, but research into the cosmic web is much broader. We very briefly summarise broader aspects of cosmic web research in Section~\ref{sec:cw-further-aspects}.

\section{Dark matter halos to galaxies}\label{sec:halos-to-galaxies}

\subsection{The formation and growth of dark matter halos}\label{sec:halo-growth}
The collapse and virialisation of dark matter halos is well understood. On large scales, where density fluctuations are small, linear perturbation theory describes the growth of the dark matter density fluctuations remarkably well. However, dark matter halos represent large density contrasts, for which linear theory does not apply. It is possible to follow the evolution of density fluctuations into the non-linear regime in a very simple way if we consider dark matter overdensities as spherical isolated regions embedded in a homogeneous expanding background. In this case, forces external to the spherical overdensities cancel out from symmetry and we can follow their evolution using the Friedmann equation in a closed model. Spherical overdensities initially expand with the background but, if they are above a certain critical density, they eventually collapse. Due to the random orbits of dark matter particles in halos and departures from pure isotropy, the collapse leads to the halo’s virialisation, rather than its formal collapse into a singularity. When this happens, the value of the spherical overdensity is around 200 times that of the background. So, in summary, we can think about dark matter halos as approximately spherical, isolated, and virialised structures with a density of around 200 times that of the background.

However, it is oversimplistic to think about a halo ``forming" at a specific time. Structure formation is a continuous process under gravity. Smaller overdensities (deeper into the non-linear regime) collapse earlier, and then merge with other collapsed structures to create larger halos. This hierarchical growth \citep{Peebles-1980} is a deeply non-linear process. Nonetheless, a simple analytic theory for the development of hierarchical structure was proposed by \cite{Press-Schechter-1974} by computing the number of collapsed objects inside regions of different sizes and final mass. Although simple, and only accurate to around 50\%, the Press and Schechter formalism completes a narrative that allows us to go from linear growth to a halo mass function in a fully analytic form. It also allows us to understand why halos are biased (i.e., their distribution does not map that of the full matter field on a 1:1 basis): peaks that sit on a background large-scale overdensity are more likely to cross the critical density threshold, and therefore undergo collapse. This {\it peak-background split} mechanism \citep{Bardeen-1986}, together with the halo mass function, then gives us analytic predictions of the large-scale clustering of halos as a function of their mass. The bottom line is that more massive halos are more biased (i.e. cluster more strongly) and began forming earlier. On large scales, this intuition also holds true for galaxies. 

None withstanding the impressive analytic progress that has been made since Press-Schechter (see e.g. \citealt{Bond-1991, Sheth-2001}), structure formation at the halo scale is now better understood using N-body dark-matter simulations. These have revealed that halos do not only grow hierarchically, via the merging of smaller halos, but also through diffuse accretion of dark matter from their (anisotropic) surroundings \citep{Fakhouri-2010}. The details of how halos grow (their accretion rate and the relative growth from diffuse accretion or mergers) depends strongly on their mass and on their environment, and will be reviewed in more detail in Section~\ref{sec:CW-halos} within the context of the larger anisotropic environment (the cosmic web). 

\subsection{Dark matter halo and galaxy growth}\label{sec:halos-galaxy-growth}
We saw in the previous section that dark matter halos grow via mergers and diffuse accretion. Considering galaxies, the simplest assumption is that dark matter and baryonic content of halos grow together. If gas formed stars with the same efficiency at all times and at all halo and stellar masses, then the star-formation histories of galaxies would map the assembly history of halos perfectly, at least for central galaxies. In that scenario, the ratio of stellar to halo mass of central galaxies would also show no dependence on halo or stellar mass. In practice, however, we know the story is more complicated than that. For example, star formation efficiency depends on the density and temperature of the gas, meaning that feedback events - such as energy injection from Active Galactic Nuclei (AGN) or supernovae - can modulate the conversion of gas into stars in ways that correlate with halo or galaxy stellar mass. How halos accrete mass also depends on their mass and environment. The relationship between halo and galaxy growth is therefore full of complexity. The {\it end} result of this coupled growth is encapsulated in the stellar-to-halo mass relation (SHMR), which is directly observable (see Section~\ref{sec:SHMR}). However, a more direct understanding of how galaxies and halos grow is hindered by the difficulty of measuring mass assembly histories directly, and therefore insight in this field is set predominantly by simulations. 

Cosmological hydrodynamical simulations, such as EAGLE \citep{Crain-2015, Schaye-2015}, IllustrisTNG \citep{Springel-2018} or Horizon-AGN \citep{Dubois-2014}, have transformed our ability to study the relationship between halo and galaxy growth. By focusing on halos that have the same mass at present day, it is possible to ask how different assembly histories modulate the properties of the galaxies they host. Studies found that the assembly history of a halo impacts on the stellar mass, star-formation history, present-day star-formation rate, kinematics, circumgalactic medium content, and morphology of galaxies (see e.g. \citealt{Romano-Diaz-2017, Montero-Dorta-2021,  Davies-2021, Cui-2021, Alarcon-2023}). The proposed physical processes by which these happen, however, are varied and have a dependence on the simulation used. We can take the example of the role of black hole growth and AGN feedback at early times. \cite{Davies-2021} carefully modified the initial conditions of certain halos in the EAGLE simulation to perform controlled experiments. They demonstrated that earlier forming halos grow more massive black holes earlier on, which increases the amount of energy injected into the galaxy, via AGN feedback, with respect to the binding energy of the baryons. This in turn expels gas from the circumgalactic medium, lowering its density and increasing its cooling time. The result is an inefficient replenishment of gas in the interstellar medium and subsequent quenching of the galaxy. In contrast, \cite{Cui-2021} used the SIMBA simulation \citep{Dave-2019} to find that the faster buildup of gas in early accreting halos leads to a higher gas density and more effective cooling, which is able to sustain higher levels of star-formation early on. Coupled with a higher Eddington ratio and the onset of a relatively mild form of AGN radiative feedback, central galaxies in early forming halos do not experience the same level of quenching as galaxies in late forming halos, and remain star-forming (and blue) to the present day. We highlight these two studies in particular as an example of how specific formulations for star formation and AGN feedback can result in different predictions to the impact of halo assembly on galaxies, but how in neither of them halo assembly could be ignored. There are other such tensions in the literature, which we see as inevitable given the theoretical uncertainty in sub-grid physics inherent to these simulations. Nonetheless, cosmological hydrodynamical simulations will continue to be the gold standard in trying to understand the impact of structure formation on galaxies.

So, what do observations tell us? The difficulty here lies in the fact that we lack a direct observational probe of halo assembly. Efforts to understand the role of structure growth on galaxies can sit in one of two (very broadly defined) camps: empirical models, or the stellar-to-halo mass relation. In the first case, we parametrise the halo assembly history and its impact on the properties of galaxies using simple models, and then compare the different models with observations. For example, \cite{Tinker-2017} use a halo occupation distribution (HOD) model to populate halos with galaxies according to the mass of the halo. They then use an abundance matching methodology to assign each galaxy a specific star formation rate (sSFR) according to the age of the halo. The abundance matching is done in a way that the rank order of sSFR in observations is preserved in the rank order of halo ages in the mocks, at fixed halo mass - i.e., the youngest halo is assigned the highest value of sSFR, the second youngest halo is assigned the second highest value of sSFR and so on. By doing this, they can test whether a model that directly (and empirically) links halo age with present-day sSFR can match observations - in this case, the fraction of quenched galaxies as a function of large-scale galaxy overdensity. They conclude that such a model can explain the dependence of quenched fractions on density for high stellar mass galaxies, but not for low stellar mass ones. The advantage of simple empirical models such as abundance matching and HOD is that they are sufficiently simple that relationships between different variables can be tested. In a follow-up paper, for example, the same authors used abundance matching to link SFR to halo assembly rate, this time focusing on star-forming galaxies. Overall they conclude that halo assembly also leaves a correlated imprint on the properties of star-forming galaxies. Simple empirical models have been deployed extensively in the literature in an attempt to quantify the impact of halo age on the properties of galaxies. Whereas the above example focused on reproducing observed trends between galaxy properties and large-scale galaxy overdensity at fixed halo mass, others focused for example on galaxy clustering as a function of stellar mass and other properties, such as colour or star formation rate. However, there is currently no consensus on the observational imprint of halo age on the properties of galaxies at fixed halo or stellar mass from such models - see \cite{Wechsler-Tinker-2018} for a review.

\subsection{The stellar-to-halo mass relation}\label{sec:SHMR}

The other camp, when it comes to inferring observationally the impact of halo age on galaxy properties, relies on measuring galaxy properties at fixed halo and stellar mass, and look for residuals that might correlate with halo age. As we saw, the SHMR encapsulates the integrated impact of halo and galaxy growth up to the time at which it is observed, and has taken centre stage in both observations and simulations, with key observations and simulations in substantial tension with one another. 

Observationally, although the SHMR is much more accessible than halo assembly histories, estimating halo masses is nonetheless challenging, especially at large radii ($> 50$ kpc). Direct ways to estimate total halo masses include weak lensing profiles and satellite kinematics. Group catalogues are able to identify galaxies that belong to a given halo, and can straightforwardly infer group stellar mass (or luminosity) above the completeness threshold of the sample. However, associating group masses to halo masses requires calibration with, e.g., weak lensing or simulations. Dynamical modelling of stars or gas typically probe only the inner regions of a halo although, more recently, the combination of stellar with HI gas kinematics has shown to be a promising way to estimate total masses to larger radii. Indirectly, one may estimate halo masses by studying the clustering of galaxies and simple galaxy-halo connection models that populate halos with galaxies according to simple prescriptions directly relating halo with stellar mass. As halo bias depends on halo mass, the observed clustering of galaxies is able to constrain the SHMR. It is also possible to combine different approaches in self-consistent ways to infer the SHMR and its evolution with redshift (e.g. \citealt{Leauthaud-2012, Shuntov-2022}).

The mean of the SHMR is reasonably well constrained in the local Universe, and there is generally good agreement on its shape and evolution, in spite of the wide variety of methods used to infer it. The SHMR has been shown to evolve weakly with redshift up to $z\approx 4$, with a steeper evolution at higher redshift \citep{Leauthaud-2012, Behroozi-2019, Shuntov-2022}. The ratio of the stellar-to-halo masses, $M_*/M_h$, has a characteristic peak at $M_h \sim 10^{12} M_\odot$, indicative of a higher net efficiency of these halos to form stars, per unit halo mass. Broadly speaking, at lower halo masses (a central-dominated regime), this efficiency is thought to be reduced by stellar feedback processes and, at higher halo masses (a satellite-dominated regime) by AGN feedback processes or environmental processes that quench or otherwise prevent satellites from forming new stars. 

The sensitivity of the SHMR to the details of halo and galaxy growth, however, comes predominantly from investigating whether central galaxies with different stellar ages follow the same SHMR (with some intrinsic scatter) or whether they are systematically offset from the mean relation. This is because, in simulations, there is good agreement that, at fixed halo mass, central galaxies with a higher stellar mass live in older halos (or, equivalently, at fixed stellar mass galaxies with a lower halo mass live in older halos - see e.g. Fig 3 of \citealt{Tojeiro-2017}). If one assumes that halo age leaves some imprint on the properties of galaxies, one might then expect those same galaxy properties to also change according to their position in the SHMR relation. This is indeed the case. However, there is tension on the nature, direction, and explanation of these residuals. For example, \cite{Scholz-Diaz-2024} measured the stellar ages, chemical abundances, morphology, and stellar angular momentum of central galaxies as a function of their position in the stellar-to-halo mass relation. Their results also show that, at fixed total mass (from dynamical modelling), central galaxies with a larger stellar mass have older stellar populations, are less rotationally supported, are more likely to have early-type morphologies, and are more metal rich (see also \citealt{Oyarzun-2022}). This result is in apparent tension with earlier observations using stacked weak lensing profiles that showed that, at fixed stellar mass, redder galaxies have a larger halo mass than blue galaxies \citep{Mandelbaum-2016}. This remains unresolved, with no clear consensus on the direction of the residuals of the SHMR with galaxy colour (taken here as a proxy for stellar age): i.e. at fixed halo mass, do galaxies with higher stellar mass have younger or older stellar populations? Whereas it is tempting to associate older halos with older stellar populations,  we recall that the impact of halo growth on the properties of galaxies depends strongly on the model used (as per the example given in the previous section), and remains uncertain even in hydrodynamical simulations. Suggested solutions include systematic errors related to the wide range of methods used to infer halo and galaxy properties, and the non-trivial inversion of the SHMR whereby stacking or averaging based on stellar mass (as is required for stacked weak lensing profiles) leads to different results from stacking or averaging based on halo mass (see e.g. \citealt{Moster-2020}).\\

In summary:\\

\begin{itemize}
\item Dark matter halos are virialised, gravitational bound structures that have decoupled from the Hubble flow. More massive halos cluster more strongly and begin assembling earlier.
\item Halos and galaxies grow hierarchically via merging and through accretion of material from their anisotropic environment.
\item Simulations indicate that older halos host more massive central galaxies, at fixed halo mass.
\item However, the impact of halo growth on the properties galaxies, at fixed halo or stellar mass, is unclear both in simulations and observations, with substantial tension in the literature. \\
\end{itemize}

A more holistic view of a galaxy’s and halo’s environment, which considers both larger scales and the anisotropic nature of structure formation, is required to understand the full impact of large scale structure on galaxies, and we turn to that next.

\section{Structure formation and the cosmic web}\label{sec:cosmic-web}

Large-scale structure of the Universe is defined as inhomogeneity present on scales that are larger than that of a galaxy. 
On these scales, from a few to hundreds of megaparsecs, matter is organised into a complex network of interconnected elongated filaments, emanating from dense compact peaks,  and sheet-like walls, forming the boundaries of large almost empty regions, called voids, in between (see Fig.~\ref{fig:figure-1}). One of the main characteristics of this matter arrangement is its hierarchical nature, manifested by the presence of structures over a large range of densities and scales.
Our current understanding of the formation and evolution of this so-called cosmic web is rooted in the anisotropic nature of gravitational collapse as the main driver of the formation of all structures in the Universe \citep{Peebles-1980} and the tidal field as the key ingredient shaping them on cosmological scales \citep{Zeldovich-1970}. It was realised, in the seminal work by \cite{Zeldovich-1970}, that the gravitational collapse of an anisotropic matter configuration proceeds in different stages, by successive collapse first along the shortest principal axis of the structure, giving rise to first the formation of sheet-like wall, followed by the collapse along the intermediate axis and formation of elongated filament, and eventually a dense compact region for the collapse along the longest axis. This picture is however not complete, as these processes occur in a hierarchical manner over multiple scales in an expanding Universe. 
All of these aspects of large-scale structure formation and evolution have been successfully unified in the cosmic web theory \citep{Bond-1996} providing a coherent framework for understanding the emergence, dynamics and evolution of the cosmic web. 
In this picture, the very existence of the cosmic web is encoded in the initial fluctuations in the primordial density field, that evolve under the effect of gravity in an expanding Universe. It is the high density peaks, their position and large-scale tidal field, that define the connecting bridges between them, forming a web dominated by filaments. 
Essentially all large N-body simulations of structure formation in a $\Lambda$CDM Universe confirm such characteristics of the cosmic matter distribution.  

This cosmic web picture was developed to explain the distribution of galaxies revealed progressively since the late 1970s by first observations of superclusters and early galaxy redshift surveys of nearby Universe \citep[e.g.][]{Einasto-1980}, showing coherent structures on scales beyond that of galaxy clusters, and detection of the first large cosmic voids \citep[e.g.][]{Kirshner-1981}. With systematic and large redshift surveys, starting with CfA Redshift Survey \citep[][]{deLapparent-1986}, followed by the 2dFGRF 
\citep[][]{Colless-2001},
SDSS \citep[][]{York-2000},
GAMA \citep[][]{Driver-2011}, 
Vipers \citep[][]{Guzzo-2014} 
or 
COSMOS \citep[][]{Scoville-2007}, 
the cosmic web, as traced by galaxies, has been successfully mapped from the nearby Universe up to redshift $z \sim 0.9$. In the near future, ongoing and upcoming surveys such as Euclid \citep[][]{Laureijs-2011,Mellier-2024} or 
PFS \citep[][]{Takada-2014},
will allow us to map, for the first time, the cosmic web using galaxies as the tracers of the underlying density field, up to redshift of about $z \sim 2$, close to the peak epoch of star formation.
At higher redshifts ($z>2$) the cosmic web is currently observationally accessible only through the tomographic reconstruction using the Lyman-$\alpha$ forest absorption by neutral hydrogen, seen in the spectra of bright background sources such as quasars or star-forming galaxies, as a tracer of the underlying density field. Such a three-dimensional reconstruction of the density field has been successfully performed with e.g. CLAMATO \citep[][]{Lee-2016} and eBOSS-Stripe 82 \citep[][]{Ravoux-2020} surveys, and will become progressively more and more accessible with ongoing and future surveys such as e.g. DESI \citep[][]{DESI-2016}, WEAVE-QSO \citep[][]{Pieri-2016}, or PFS.

The importance of the cosmic web stems from the fact that it naturally defines environment in which galaxies and galaxy halos form and evolve. Moreover, the detailed characteristics of the cosmic web and its principal structural components, nodes, filaments, walls and voids, are sensitive to cosmological parameters, but are also impacted by processes involved in galaxy formation and evolution. Studying the properties of cosmic web can therefore help constraining both cosmology and galactic formation and evolution physics. However it is not trivial, in practice, to classify a galaxy (or halo) density field into distinct cosmic web environments. Several different approaches have been developed to do this, e.g. identifying topological features in galaxy density fields (e.g. DisPerSE, \citealt{Sousbie-2011}), computing the local geometry on the tidal field (e.g. \citealt{Hahn-2007a}), or using the full phase-space information of tracers (e.g. ORIGAMI, \citealt{Falck-2012}). Most methods agree that voids and nodes are associated with the lower and highest density regions respectively, but the comparison across filament and wall environments is more nuanced and requires more care (see \citealt{Libeskind-2018}). In this chapter we do not specify the type of cosmic web estimator when referring to different studies, but note its potential importance in comparing and contrasting different results and resolving potential tensions. Whilst work has been done on a thorough comparison of different estimators on an N-body dark matter simulation \citep{Libeskind-2018}, the impact on galaxy properties is yet to be fully understood.   \\

In summary:\\

\begin{itemize}
\item The matter, including halos and galaxies, is on large scales organised into nodes, filaments, walls and voids, forming the so-called cosmic web.
\item The cosmic web originates from the initial fluctuations in the primordial density field, enhanced by anisotropic gravitational collapse at later times.
\item There are various methods to identify the different cosmic web environments from a galaxy or halo density field. 
\end{itemize}

\section{The impact of the cosmic web on halos and galaxies}\label{sec:CW-halos-galaxies}

\subsection{The impact of the cosmic web on the properties of dark matter halos}\label{sec:CW-halos}
According to the simple description presented in Section~\ref{sec:halo-growth}, the formation and growth of dark matter halos depend entirely on their mass. However, simulations also show that halo properties depend on their environment or halo formation time, at fixed halo mass. It became clear, for example, that older halos cluster more strongly (\citealt{Gao-2005}), a phenomenon that became known as {\it halo assembly bias}. In fact, the large-scale clustering of halos was shown to have a secondary dependence on several other halo properties, such as halo concentration, shape, and velocity anisotropy (e.g. \citealt{Faltenbacher-2010}). At high mass, some degree of assembly bias seems unavoidable due to strong correlation between a halo's Lagrangian patch (defined by the region in the initial conditions from which the halo originates) and its environment, and is indeed predicted by simple structure formation models based, e.g., on peak theory \citep{Dalal-2008}, although with a different direction: old, or more concentrated, halos cluster more weakly. At low halo mass, however, the cosmic web is now widely seen as responsible for modulating the growth and properties of dark matter halos and, consequently, for a significant proportion of the halo assembly bias signal in general.

The dominant impact of the cosmic web on halos comes through their accretion rate. Early studies into halo assembly bias established that low mass halos in the vicinity of massive halos stop accreting, providing a natural explanation for assembly bias at low halo mass. But the physical processes at play were not immediately clear. \cite{Hahn-2009} showed that the accretion rate of low mass halos correlated strongly with the tidal field, due to the sheared flow around low mass halos. Working in zoom-in hydrodynamical simulations, \cite{Borzyszkowski-2017} showed the complex behaviour of the flow of matter around "stalled" and "accreting" halos due to sheared velocity fields, and its link with cosmic web environment (see panels D and E of Fig.~\ref{fig:figure-1}). Stalled halos (panel D) - not undergoing significant accretion - were found to reside in filaments thicker that the virial radius of the halo. Within the rest frame of the halo, matter flow along the filament recedes from the halo, preventing accretion in this direction. In stalled halos, therefore, accretion is only possible tangentially to the filament. In contrast, accreting halos are found at the nodes of thin filaments (panel E), that continue to enable accretion more or less isotropically (see also \citealt{Ganeshaiah-Veena-2018}). These thin filaments are not sufficiently massive to create strong anisotropic tides, and are distinct from the large filaments traced by galaxies in observations. In other words, accreting halos might be present in underdense regions (such as voids) in current observations. A link between stalled halos and clustering amplitude can be inferred by the fact that stalling happens preferentially in dense environments, but a direct link between large-scale halo bias and the anisotropy of the tidal environment of a halo was clearly shown in \cite{Paranjape-2018} (see also \citealt{Balaguera-Antolinez-2024}).

The different accretion modes in stalled and accreting halos has consequences for the dynamics of dark matter particles, subhalos, and satellite galaxies. Stalled halos (associated with thick filaments) have preferentially tangential orbits, whereas accreting halos (far from the influence of a massive neighbour or filament) are able to accrete radially. This leads to a noticeable difference in the velocity anisotropy of dark matter particles, subhalos and satellite galaxies according to cosmic web environment (\citealt{Borzyszkowski-2017,Garaldi-2018}), explaining the strong bias segregation with velocity anisotropy seen in \cite{Faltenbacher-2010}. 

Simulations also show that stalled halos are more concentrated and store a larger fraction of their mass in the central subhalo - i.e., they have a lower substructure fraction (\citealt{Wang-2011, Borzyszkowski-2017}). This led \cite{Lim-2015} to propose the ratio of the stellar mass of the central galaxy to the total halo mass ($M_*/M_h$) as an observational proxy for substructure fraction and, therefore, halo age. They find that, at fixed halo mass, central galaxies with a higher $M_*/M_h$ are redder, which is interpreted explicitly as being due to residing in older halos. However, given the theoretical uncertainty in mapping halo assembly to galaxy properties (see Section~\ref{sec:halo-growth}), studying the $M_*/M_h$ in the context of the cosmic web can help clarify the link between halo assembly and galaxy properties. This was done in \cite{Tojeiro-2017} who found that, at fixed low halo masses, central galaxies in knots and filaments have a higher $M_*/M_h$ than those in voids and walls, lending observational strength to the argument that the cosmic web modulates halo age (i.e., galaxies in knots and filaments reside in older - or stalled - halos). At high halo mass, the trend disappears, with mild evidence that it is reversed, in agreement with theoretical predictions. A combination of substructure fraction and galaxy/halo kinematics may also be combined into a useful probe of halo assembly. A link between $M_*/M_h$, stellar/gas kinematic misalignment and the cosmic web environment was proposed by \cite{Duckworth-2019}, working on integral field spectroscopy data from the Mapping Nearby Galaxies at APO (MaNGA, \citealt{Bundy-2015}), but they found no conclusive link between the three. They conclude that kinematic misalignment in galaxies is not driven by external factors and leave no significant imprint within 2.5 effective radii (as probed by MaNGA).  Tidal field anisotropies are expected to leave strong kinematic signatures only to much larger radii \citep{Romano-Diaz-2017}, so better and larger samples will be required. Current and future surveys might also be able to test the impact of the cosmic web on the growth of dark matter halos at larger radii via satellite dynamics, providing a new platform for testing structure formation and galaxy evolution models. 

Although halo assembly bias is well established in simulations, whether and how halo assembly bias propagates to the properties of galaxies (often referred to {\it galaxy assembly bias}) remains an open question. It is also a critically important question in observational cosmology studies due to the need to interpret small and large scale galaxy clustering within the context of the underlying matter density field. Outside of the specific context of the cosmic web (and therefore of the scope of this article), quantifying and qualifying galaxy assembly bias in observations and simulations has therefore been a very active area of research - see \cite{Wechsler-Tinker-2018} for a review.\\

In summary:\\
\begin{itemize}
\item Simulations show that the cosmic web impacts on the properties of dark matter halos, at fixed halo mass: halos near or within large filaments are older (they have a lower accretion rate), have a lower substructure fraction, and have preferentially tangential orbits. 
\item In contrast, halos far from the strong tidal effects induced by the cosmic web accrete more strongly, have a larger amount of substructure, and have more radially-dominated orbits.
\item The dependence of halo properties on cosmic web environment can explain halo assembly bias at low halo mass. 
\end{itemize}

\subsection{The impact of the cosmic web on the scalar properties of galaxies}\label{sec:CW-galaxies-scalar}

Being prototypal examples of the environment, large galaxy groups and clusters, typically residing in high-density regions of the Universe, have been extensively studied in the context of their impact on galaxy properties. These studies revealed today’s well-known relations involving the local density, such as mass-density \citep[][]{Dressler-1980}, morphology-density \citep[][]{Dressler-1980}, and colour-density \citep[e.g.][]{Balogh-2004} or SFR-density \citep[e.g.][]{Hashimoto-1998} relations. Conversely, studies involving other components of the cosmic web, i.e. not only nodes, but also filaments, walls and voids, have only recently started to gain full attention. 
There are several reasons for this. 
One of them is the only relatively recent development of multi-scale algorithms allowing the reconstruction of the cosmic web from discrete data sets, such as galaxy catalogs. In addition, observationally, such a reconstruction in 3D is in general very challenging, as it requires both high spatial sampling of galaxies and accurate measurements of their redshifts.  
%
%
Finally, the signal of the impact of the anisotropic environment on galaxy properties is expected to be difficult to extract, as it is a second order effect than needs to be first disentangled from the effect of the density, an isotropic measure of environment\footnote{We will use the term density loosely to indicate any isotropic environmental measure without specifying the scale nor the method used for its reconstruction. Detailed discussion on the impact of the density on the cosmic web-related measurements can be found in e.g. \cite{Kraljic-2018}.}. Different methods are adopted in the literature to isolate the effect of the anisotropic cosmic web environment from the one driven by the local density, but the main philosophy is always the same. 
This consist of keeping the density distribution of a given population of galaxies unchanged while breaking the underlying dependence on the cosmic web. This is often done by shuffling galaxies within the regions of same local density or simply performing analysis in different density bins. Any remaining residual trend with the large-scale anisotropic environment can then be interpreted as being a signature of the impact of cosmic web beyond the effect of the density alone. 
A major obvious difficulty of this exercise is the definition of the density itself, which depends not only on the adopted method but also on the scale on which it is computed, probing potentially different physical processes and time scales over which these operate. Typically, the larger the scale, the more integrated the effect of the galaxy’s environment is while, at small scales, the density is more correlated with recent stochastic events. 
Therefore, in order to infer the statistical impact of the large-scale structure on galaxies, scales large enough to average out these recent local events should be considered, i.e. larger than the mean inter-galactic separation \citep[for a detailed discussion on this topic, see e.g.][]{Kraljic-2018}. Consequently, it is not immediately clear how to interpret and directly compare results based on different methods. 

All these aspects contribute to discrepancies in drawn conclusions on the impact of explicitly the cosmic web environment on different galaxy (and/or halo) properties. To give a few examples, among earlier studies, \cite{Eardley-2015} found 
that the observed significant variation of the luminosity function of galaxies in the Universe can be entirely attributed to the indirect local-density dependence.
More recently, \cite{O'Kane-2024} similarly found that once matched in local density, star formation activity and morphology of filament and field population of observed galaxies are indistinguishable. However, their choice of density estimator traces small scales ($\lesssim$ 1 Mpc) where it is expected not to find any difference with the large-scale environment \citep[see][]{Kraljic-2018}.  

In the remainder of this section we summarize observational findings that provide evidence of the influence of the cosmic web on different galaxy properties. We focus on large-scale components of the cosmic web, as traced by the distribution of galaxies in both large galaxy surveys and hydrodynamic simulations, and compare with theoretical predictions and expectations.

\paragraph{Stellar mass} One of the most fundamental integrated properties of galaxies is their stellar mass. 
Observational findings of massive galaxies residing preferentially in dense environments, such as groups and clusters, can be theoretically understood through the biased mass function close to large-scale density enhancement of the underlying dark matter density field favouring an earlier collapse of bound structures, resulting in an overabundance of massive halos in dense environments. 
The anisotropic large-scale environment acts as a long-wavelength density mode on top of the local overdensity that allows halos to pass the turnaround threshold. As a result, a shift of the halo mass distribution towards larger masses is expected in the vicinity of large-scale structure. Observationally, such a trend has been indeed identified with more massive galaxies being preferentially located closer to large-scale filaments of the cosmic web compared to their lower mass counterparts, at fixed local overdensity, in the low-redshift Universe  \citep[e.g.][]{Chen-2017,Kraljic-2018,Winkel-2021}, up to redshifts of $\sim$ 0.9 \citep[][]{Malavasi-2017,Laigle-2018}. 
These trends are not limited to the filamentary network alone, they have also been identified for a lower density large-scale environment such as walls \citep[][]{Kraljic-2018,Winkel-2021}. 
Qualitatively similar stellar mass gradients are recovered in (literally all) large-scale hydrodynamic cosmological simulations \citep[e.g.][]{Kraljic-2018,Laigle-2018,Bulichi-2024}.

\paragraph{Colour/sSFR}
Galaxies grow in mass through the conversion of gas into stars and through mergers. The process of star formation is therefore closely connected to that of galaxy evolution. As galaxies evolve within the cosmic web, it is crucial to understand the role this anisotropic environment plays, if any, in regulating the star formation activity beyond that of the stellar mass. Hint of such an impact has been revealed by observational studies in the local and higher-$z$ Universe ($z \lesssim 0.9$) finding that  passive galaxies tend to be located closer to the spine of filaments \citep[e.g.][]{Chen-2017,Kuutma-2017,Kraljic-2018,Laigle-2018,Winkel-2021} and closer to walls \citep[][]{Kraljic-2018} compared to star-forming galaxies of same stellar mass. Qualitatively similar sSFR gradients with respect to filaments and walls were recovered by different large-scale hydrodynamic simulations \citep[][]{Kraljic-2018,Laigle-2018,Hasan-2023,Bulichi-2024}. 
Additional clues on the impact of filaments on the star formation activity of galaxies were gained by measuring the passive (red or quenched) fraction, i.e. the number of passive galaxies over the total population, in the low-$z$ Universe showing a clear trend with both distances to nodes of the cosmic web, but also to filaments \citep[e.g.][]{Kraljic-2018}. As expected from the colour-density (or sSFR-density) relation known from studies of galaxy clusters, the fraction of passive galaxies was found to increase with decreasing distance from the nodes of the cosmic web. However, this fraction was found to increase, to a lesser extent, also when closing in on filaments at fixed distance from nodes. This clearly suggests that galaxies are impacted by the large-scale environment even before they fall into groups and clusters, a phenomenon known as ‘pre-processing’ (see e.g. \citealt{Kuchner-2022}). This trend of increasing quenched fraction with decreasing distance from filaments of the cosmic web was recovered in hydrodynamic cosmological simulation SIMBA at redshifts $z \le 1$ and found to become stronger with decreasing redshift \citep[][]{Bulichi-2024}.

While it is expected that the large-scale anisotropic tidal field impacts the assembly history of dark matter halos (see Section~\ref{sec:CW-halos})
it is not obvious how to directly connect their mass accretion rate to properties of galaxies, and predict, from the first principles, its impact on e.g. star formation or colour at fixed mass and local density (see Section~\ref{sec:halos-galaxy-growth}). Therefore, the exact cause of decreased star formation activity of galaxies in the vicinity of filaments is still debated. Several interpretations were proposed in the literature such as the cosmic web detachment, where galaxies are detached from filamentary accretion of cold gas at shell-crossing on intergalactic scales \citep[][]{Aragon-Calvo-2019}, the suppression of star formation close to filaments caused by high angular momentum supply at the vorticity-rich filaments’ edges leading to a less efficient transfer of gas from outer to inner regions of halos near filaments \citep[][]{Song-2021}, enhanced AGN feedback, leading to stronger quenching, due to stronger accretion on halos close to large-scale filaments \citep[][]{Kraljic-2018,Musso-2018}, suppressed accretion of external gas and gas loss via ram pressure for halos having recently crossed walls and filaments within them \citep[][]{Thompson-2023} or 
shock heating of the gas as a dominant effect reducing star formation near filaments to which an extra suppression is added by the effect of AGN feedback, in particular at low redshift \citep[][]{Bulichi-2024}, to mention a few.
All of these processes should also impact other galaxy properties beyond their star formation rate, related to both their stellar and gas content.

\paragraph{Stellar properties}
Indeed, a detailed study of a sample of central galaxies in the low-$z$ Universe ($z \lesssim 0.2$, using SDSS data) revealed that while the specific star formation rate of galaxies decreases with decreasing distances from nodes, filaments and walls, their age, stellar metallicity and element abundance ratio [$\alpha$/Fe] increase (after controlling for stellar and halo mass) \citep[][]{Winkel-2021}. 
A possible interpretation is the removal of the metal-poor gas in the outskirts of galaxies in the vicinity of individual cosmic web components. The metal-rich gas in the central regions of galaxies could continue forming stars rich in metals. A direct consequence of such a scenario would be the increased gas-phase metallicity of galaxies enhanced by the reduced access to the pristine gas from the cosmic web that could dilute the metallicity of the gas in the ISM of galaxies.

The enhanced stellar metallicity for central galaxies near filaments seems to be in contradiction with measurements in hydrodynamic simulation SIMBA \citep[][]{Bulichi-2024}. While galaxies, both centrals and satellites, are indeed found to be more metal-enriched as their distance from filaments decreases, when accounting for the mass-metallicity relation, no trend is found for centrals. However, a strong trend persists for satellite galaxies, showing a significant enhancement of metallicity close to filaments. This suggests that strong metallicity gradient with the distance from filaments is for centrals essentially driven by the stellar mass, with more massive (and more metal-rich) galaxies being preferentially located near filaments. Nevertheless, given that central galaxies in this simulation continue to show a strong suppression of sSFR even after accounting for the stellar mass dependence, it was suggested that the fundamental metallicity relation (FMR; Mannucci et al. 2010) is predicted to depend on the exact location within the cosmic web \citep[][]{Bulichi-2024}.

\paragraph{Gas properties}
Observational findings for low redshift star-forming galaxies show that their proximity to a node and, to a lesser extent, to a filament modulates their gas-phase metallicity \citep[][]{Donnan-2022}. Galaxies closer to nodes and filaments display higher chemical enrichment compared to those further away, an effect that was shown to be independent of their stellar mass and over-density. This observational finding of the scatter in the galaxy stellar mass-gas metallicity relation being correlated with the distance to nodes and filaments is in qualitative agreement with cosmological hydrodynamic simulation TNG300 \citep[][]{Springel-2018} providing an explanation through the combination of halo assembly bias and gas supply in the vicinity of the cosmic web, nodes in particular. In this interpretation, galaxies near nodes have limited access to metal-poor gas that could dilute the metallicity of already existing gas and, at the same time, their star formation history was more active in the past, meaning that their interstellar medium have been enriched with more metals compared to galaxies of the same mass in different cosmic web environments \citep[][]{Donnan-2022}. 
This seems to be in contradiction with results on the stellar metallicity in the SIMBA simulation mentioned in the previous subsection. It is however difficult to draw a firm conclusion from this comparison, as in the work of \cite{Donnan-2022} the total population of galaxies, both centrals and satellites, was considered. Another possibly important difference is their focus on star forming galaxies alone, while no condition on star formation activity was imposed in the work of \cite{Bulichi-2024}. 
This nevertheless highlights the difficulties that current surveys need to deal with, that is being able to reliably classify galaxies into centrals and satellites.  

Constituting a major part of the gas content of galaxies and representing the dense gas reservoir for eventual star formation, atomic neutral hydrogen, HI, is a crucial ingredient of galaxies' baryon cycle. 
Focusing on massive galaxies, with stellar mass above $10^{11} M_{\odot}$, from 6dF Galaxy survey and HIPASS, \cite{Kleiner-2017} found that their HI fraction is higher near filaments compared to the control sample of galaxies at same projected density and minimum distance from filaments. For lower mass galaxies, no statistically significant difference was detected. 
This was interpreted as evidence for replenishment of HI gas through accretion of cold gas from filaments in massive galaxies, supporting the cold mode accretion scenario where galaxies with a large gravitational potential are capable of drawing gas from large-scale filaments in their vicinity. 
Conversely, the analysis of galaxies from the ALFALFA HI survey revealed that at fixed local density and stellar mass, and after removing group galaxies from the sample, late-type galaxies in the stellar mass range $8.5 < \log(M_{\star}/M_{\odot}) < 10.5$ show increased HI deficiency with decreasing distance from filaments, suggesting a cut off from the cold gas supply for galaxies in the neighbourhood of large-scale filaments \citep[][]{Crone_Odekon-2018}. 
This result is in qualitative agreement with trends from hydrodynamic cosmological simulation SIMBA \citep[][]{Bulichi-2024}, where central galaxies show suppressed HI gas content close to filaments, with suppression being strongest at $z=0$. Satellite galaxies show a comparable trend except at very small distances from the core of filaments where the HI fraction increases at $z=0$, but is most likely due to the HI assignment scheme adopted in the simulation that might associate the HI of the group environment to satellites. However, when accounting for the stellar mass dependence of the HI gas mass fraction, central galaxies exhibit reverse trend, with enhanced HI fraction close to filaments. 

\paragraph{Beyond the local Universe}
Most of the observational findings on the impact of the cosmic web on different galaxy properties are restricted to the low-redshift Universe, where existing redshift surveys offer high enough sampling to allow for studying the co-evolution of galaxies and their large-scale anisotropic environment. In the near future, spectroscopic surveys such as Euclid and PFS, will allow us to extend this analysis to higher redshifts, up to $z\sim2$. However, hydrodynamic cosmological simulations already today provide predictions
for these cosmic epochs. The suppression of sSFR of galaxies, at fixed stellar mass, near filaments seen at low redshifts is predicted to be already in place by $z=1$ for both centrals and satellites, while at even higher redshifts ($z\sim2$) no strong trends with a distance from filaments is expected \citep[][]{Hasan-2023,Bulichi-2024}. Similarly, no obvious trend is seen in hydrodynamic simulation at $z=2$ for the quenched fraction, however at $z \leq 1$ the fraction of quenched galaxies, both centrals and satellites, is found to increase with decreasing distance from filaments, with a strength that increases with decreasing redshift \citep[][]{Bulichi-2024}.
Interestingly, while central galaxies continue to show no strong trend for stellar metallicity once the stellar mass dependence is accounted for, for satellite galaxies on the contrary a strong enhancement of metallicity in the vicinity of filaments is found to persist up to $z=2$ \citep[][]{Bulichi-2024}.  
Predictions concerning the gas content of galaxies point towards a suppressed HI fraction near filaments already at $z\sim2$, with an amplitude and spatial extent around the filaments that increase with time \citep[][]{Bulichi-2024}, 
in a qualitative agreement with the decreased total gas mass fraction around filaments \citep[][]{Hasan-2023}. 
When accounting for the stellar mass dependence of the HI fraction, central galaxies show enhanced HI content in the vicinity of filaments with an amplitude stronger at $z=2$ compared to the local measurement. Satellites, on the other hand, continue to show a strong suppression of HI content at high redshifts ($z=1-2$) beyond the effect driven by stellar mass. \\

In summary:\\

\begin{itemize}
\item Observations in the low-redshift ($z \lesssim 0.3$) Universe provide evidence that the anisotropic large-scale environment impacts at least some of galaxy properties, beyond their stellar mass and underlying density. 
\item sSFR, metallicity and HI gas content of galaxies are in particular found to be sensitive to their position within the cosmic web. 
\item Galaxies that are located closer to filaments of the cosmic web are not only more massive, but tend also to have lower sSFR and higher (stellar and gas-phase) metallicity. 
\item The trends with the HI fraction are debated, potentially due to selection biases. However they show a clear dependence on the large-scale environment. 
\item Hydrodynamic simulations provide overall support for these findings. However, more detailed analysis is needed to understand some of the remaining discrepancies. 
\end{itemize}

\subsection{The impact of the cosmic web on the spin alignment of galaxies}\label{sec:CW-galaxies-spin}

Beyond scalar properties, the orientation of angular momentum of galaxies and their shape also show a dependence on their large-scale environment. 
In the local Universe, there is a good agreement between different observations of elliptical/S0 galaxies finding a preferential orthogonal orientation of their angular momentum (or minor axis) with respect to their host filaments \citep[e.g.][]{Tempel-2013a}, in line with results of galaxies' intrinsic shape measurements \citep[e.g.][]{Joachimi-2011}. 
A worse agreement seems to exist for disc galaxies.
While some early studies found preferentially parallel orientation for spirals \citep[e.g.][]{Tempel-2013a}, Scd types \citep[][]{Hirv-2017}, or both red and blue galaxies \citep[][]{Zhang-2013}, others reported a tendency for a perpendicular orientation for spirals \citep[e.g.][]{Lee-2007} and Sab galaxies \citep[][]{Hirv-2017}, or lack of any evidence for a clear signal \citep[e.g.][]{Pahwa-2016}. More recent studies employing high-quality stellar and gas kinematics from integral field spectroscopy surveys such as MaNGa \citep[][]{Bundy-2015} and SAMI \citep[][]{Croom-2012}, across a range of large-scale environments, seem to support a picture in which late-type or low-mass galaxies have their angular momentum preferentially aligned with their nearest filament, while the early-type/S0 or massive galaxies display an orthogonal orientation \citep[][]{Welker-2020,Kraljic-2021}. 
The primary parameter driving the signal seems to be the mass of the bulge, as galaxies with lower bulge masses are found to have their angular moment parallel to the closest filament, while galaxies with higher bulge masses show a tendency for perpendicular orientation \citep[][]{Barsanti-2022}. 

From the theoretical perspective, an impact of LSS on spin alignment of galaxies is expected, at least to a certain degree. 
As the cosmic web defines a preferred direction of the matter infall and generates an anisotropic tidal field, it is expected that it impacts also on the internal dynamical properties of forming galaxies. According to tidal torque theory (TTT) in the standard paradigm of galaxy formation, protogalaxies are thought to acquire their angular momentum, in the linear regime, due to the misalignment between the inertia tensor of the protogalaxy and the tidal tensor exerted by the surrounding matter distribution \citep[][]{Peebles-1969,Doroshkevich-1970,White-1984}.
It was later shown \citep[][]{Codis-2015}, that taking into account the local preferred direction for both the tidal tensor and the inertia tensor of forming protogalaxies, induced by the presence of a typical filament in their vicinity, can explain their spin distribution. 
Indeed, the resulting spin field near filaments 
has a specific transverse and longitudinal geometry (reflection-symmetric with respect to the saddle point defining the filament), 
%
%
with the mean spin field parallel to the filament axis in the plane of the (filament-type) saddle point, and becoming azimuthal away from it. This ``constrained" or ``conditional" TTT also predicts that less massive halos (and possibly also galaxies) should have their spin parallel to the filament, while more massive ones are expected to have their spin in the perpendicular direction. 
This is in agreement with findings for halos in cosmological N-body simulations \citep[e.g.][]{Aragon-Calvo-2007,Hahn-2007,Codis-2012,Trowland-2013}.
State-of-the-art hydrodynamic cosmological simulations seem to suggest that galaxies also retain a memory of their spin orientation acquisition within the anisotropic cosmic web environment \citep[e.g.][]{Dubois-2014,Codis-2018a,Wang-2018}, even though the stellar mass dependence of the signal is debated. Some studies confirmed the transition from a parallel orientation with respect to filaments for low mass galaxies to a perpendicular one for galaxies at high mass end \citep[e.g.][]{Dubois-2014,Codis-2018a,Kraljic-2020a}, and analogously with respect to walls 
\citep[e.g.][]{Codis-2018a,Kraljic-2020a}. However, there are also studies finding a preferential perpendicular orientation between the spin of galaxies and the direction of their neighbouring filament, regardless of their stellar mass \citep[][]{Ganeshaiah-Veena-2019,Krolewski-2019}. These contradictory results can be reconciled by considering the multi-scale nature of the cosmic web filaments. In the conditional TTT \citep[][]{Codis-2015}, the transition mass for the spin flip depends on the underlying large-scale density such that the larger the density, the higher the transition mass. Conditional TTT predicts that if the mass of galaxy (or halo) is below this transition mass, then its spin tends to be aligned with the underlying filament, while it is perpendicular if the mass is above the transition mass. Therefore, at fixed mass, galaxies in low-density (or thinner) filaments will preferentially show perpendicular spin orientation, as in \cite{Ganeshaiah-Veena-2019}. This line of argument is supported by findings of stronger impact of large scale tides on the galaxy spin orientation in denser filaments \citep[][]{Kraljic-2020a}. 

A qualitative understanding of the mass-dependent spin flip can be reached by considering the dynamics of matter on large cosmological scales. The formation of a filament at the intersection of collapsing walls generates a vorticity field aligned with its axis and with typically quadrupolar (filament-type saddle point-reflection symmetric) geometry. The first generation of galaxies acquiring their spin from the accretion within quadrants of given orientation have therefore their spin typically aligned with the axis of filaments. At  the later stage, galaxies and halos flow towards the nodes as filaments collapse, merge, grow in mass and convert their orbital angular momentum into spin perpendicular to the direction of the filaments axis \citep[][]{Codis-2012,Welker-2014}.
The role of mergers in spin flip of massive galaxies from parallel to perpendicular orientation was recently questioned by \cite{Lee-2022} finding only a very weak, if any, dependence of halo spin alignments on recent mergers. It was instead proposed, by \cite{Moon-2024}, that the spin transition is driven by the misalignment between the initial tidal field and inertia tensor of a protogalaxy, superseding in strength the mass-dependent transition. This purely initial conditions driven origin of galaxy spin transition was suggested to explain also other types of spin-filament alignments, e.g. the morphology-dependent spin orientation found in observations and recovered in hydrodynamic simulations for the stellar kinematics used as proxy for galaxy morphology \citep[][]{Codis-2018a}.
These two pictures may not necessarily be mutually exclusive given that the angular momentum and orbital parameters of mergers are sensitive to changes in the initial conditions of the Lagrangian region from which a galaxy originates \citep[e.g.][]{Cadiou-2022}. 

Ongoing and future surveys such as Euclid, JWST or PFS will provide additional clues on the spin alignment of galaxies by allowing us to extend the existing measurement to higher redshifts ($z\gtrsim1$), where the signal is expected to be stronger, as suggested by hydrodynamic simulations \citep[e.g.][]{Codis-2018a,Kraljic-2020a}.
Understanding the origin and evolution of galaxies' angular momentum and the role played by the large-scale cosmic environment is of great interest in galaxy formation theory, e.g. through its connection to the galaxy morphology and therefore the so-called Hubble sequence \citep[e.g.][]{Hubble-1926,Roberts-1994}. But it also crucial for the weak lensing science where intrinsic alignments are significant contaminants of the measured signal \citep[e.g.][]{Kirk-2010}. And finally, as the 
transition mass for the halo spin orientation seems to be sensitive to the the dark energy model \citep[][]{Lee2020a} and total neutrino mass \citep[][]{Lee2020b}, the intrinsic spin-shear alignment is potentially a powerful complementary probe of cosmological models.\\

In summary: \\

\begin{itemize}
\item As expected by the tidal torque theory and its modifications, the orientation of angular momentum and shape of observed galaxies depend on galaxy position within the cosmic web. 
\item In the low-$z$ Universe, massive and/or early-type galaxies tend to have their angular momentum perpendicular to the direction of the neighbouring filament, while for low mass and/or late-type galaxies, their spin is found to be aligned with their closest filament. 
\item These finding are in qualitative agreement with cosmological hydrodynamic simulations.
\end{itemize}

\section{Further aspects of the cosmic web}
\label{sec:cw-further-aspects}

Throughout this article we focused very specifically on large filaments and nodes as traced by galaxies, but research into the cosmic web is much broader than what we could cover here. In this section we highlight some (though not all) of those aspects.

\subsection{The gas content of filaments}

 The diffuse baryonic content of the cosmic web, and particularly filaments, has received substantial attention in the literature. Simulations show that 30-40\% of the baryons in the local Universe exist in the form of a diffuse, warm ($10^5<T<10^7$ K), low-density gas residing in these large filaments \citep{Dave-2001}. This warm-hot intergalactic medium (WHIM) is widely considered as a solution to the “missing-baryons” problem, and efforts to capture these missing baryons in the filamentary WHIM have now been successful in, for example, absorption in X-ray spectra of background quasars \citep{Nicastro-2018} and the stacked thermal Sunyaev-Zel’dovich effect across pairs of massive galaxies \citep{deGraaff-2019}. But the complex portrait of the gas phases in different cosmic web environments in simulations is only now starting to be revealed \citep{Martizzi-2019, Galarra-Espinosa-2021}. Whereas the bulk of the diffuse gas sits in the WHIM regime, simulations also show cold components at a range of densities depending on the properties of filaments, highlighting the role of different filaments in fuelling star formation in galaxies \citep{Galarra-Espinosa-2021}.  The cold component of the filamentary diffuse gas is also hard to detect. Studies of the circumgalactic medium of galaxies have been successful in mapping the diffuse gas around galaxies up to 10 Mpc using MgII absorbers in the spectra of background quasars (e.g. \citealt{Perez-Rafols-2015}). Several studies have also revealed the anisotropic distribution of these absorbers within the virial radius of the halo, potentially distinguishing between outflows and inflow material from filaments onto the halo. Current and forthcoming surveys such as DESI, which provide a dense population of background quasars, have the potential to enable a link between the circumgalactic medium and filament scales of diffuse cold gas, as seen in absorption. At high-redshift, most of the baryonic content is in the form of diffuse gas and observationally accessible in absorption via the Lyman-alpha forest, which allows a direct and striking 3D reconstruction of the cosmic web (see \citealt{Horowitz-2022} for a reconstruction at $z\approx2.3$).\\

\subsection{The cosmic web connectivity}
The cosmic web filaments define preferred direction of the flow of matter towards high density nodes, where galaxy clusters and massive galaxy groups typically reside. 
The number of globally connected filaments to a given node, i.e. the connectivity, can therefore provide interesting clues about the assembly history of galaxies by controlling their mass and angular momentum accretion. 
Theoretically, the connectivity of cosmic fields can be predicted from first principles \citep[][]{Codis-2018b}, in agreement with measurements in cosmological simulations. When applied to astrophysical objects of interest, such as halos, it was found, as expected, that the mean connectivity increases with increasing mass \citep[see e.g.][]{Codis-2018b, Galarraga-Espinosa-2024}. Similar trend with mass was found for galaxies, both in low-$z$ observations and hydrodynamic simulation, such that more massive galaxies tend to be connected to more filaments compared to their lower mass counterparts \citep[][]{Kraljic-2020b}. Interestingly, star formation activity and morphology of galaxies also show a clear trend with the connectivity beyond that of stellar mass, such that galaxies with higher connectivity tend to have lower sSFR and tend to be less rotationally supported, or more likely to have early-type morphology, compared to galaxies with lower connectivity. Simulations also show that connectivity increases toward higher redshifts \citep{Galarraga-Espinosa-2024}. Upcoming and ongoing spectroscopic surveys such as PFS, MOONS or Euclid will allow us to extend the measurements of connectivity and its impact on galaxy properties to higher redshifts ($1\lesssim z \lesssim 2$).

\subsection{The dependence on cosmology}
According to expectations, the evolution of cosmic connectivity was also shown to depend on cosmology \citep[][]{Codis-2018b}. The connectivity is a robust quantity, given that it reflects the underlying topology of the field, and therefore it may prove to be an interesting probe of cosmological models beyond $\Lambda$CDM. Recently, the connectivity of the cosmic web was explored as a potential probe of alternative models of gravity \citep[][]{Boldrini-2024}, dark energy and neutrino masses \citep[][]{WST-2024}. 
Other properties of the cosmic web, beyond the connectivity, can be also used to put constraints on cosmology. 
Among them, exclusion zones present in the cross-correlations of critical points of the density field, such as peak-void, peak-wall, filament-wall, filament-void, were found to behave as standard rulers and therefore represent a promising tool for constraining cosmological parameters of dark matter and dark energy \citep[][]{Shim-2024}.
On the scales of individual DM halos, it was found that while their shape can be used as a cosmological probe of the nature of gravity \citep[][]{Koskas-2024}, their mass accretion rates seem to be sensitive to the nature of dark matter, and potentially distinguish between the cold and warm dark matter models \citep[e.g.][]{Dhoke-2021}. 

\subsection{Voids and clusters}
Finally, let us note that in this article we purposely avoided to talk about voids and clusters, highly non-linear
important structural and dynamical components of cosmic web, as they are covered in some detail elsewhere in this volume. Therefore we only briefly mention some of the reason why studying the cosmic voids and galaxy clusters is important. Firstly, clusters and voids, these respectively compact overdense and large underdense regions of the Universe, are integral parts of its LSS. Hence, in order to fully understand the formation and dynamics of the cosmic web, we need to understand the evolution and properties of both clusters and voids, and the impact on galaxies as they transition between all cosmic web environments. Secondly, clusters and voids are both unique and interesting regions of the Universe for studies of the impact of environment on the formation and evolution of galaxies. On the one hand, massive galaxy clusters, assembling about 4\% of the mass in the Universe and typically located near the intersection of cosmic filaments, are high density regions where galaxies and galaxies groups undergo transformation via interactions with one another and with the gas in between (see e.g. \citealt{Kuchner-2022}). On the other hand, voids, low density regions accounting for about 95\% of the total volume of the Universe, represent a pristine environment, ideal for studying the evolution of galaxies in the absence of external processes.
Finally, both clusters and voids contain valuable information about the underlying cosmological model, allowing to constrain e.g. the dark energy equation of state, total mass of the neutrinos or different modified gravity scenarios \citep[see e.g.][]{Allen-2011,Pisani-2019}.

\section{Conclusion}

Large scale structure and the cosmic web leave an imprint on the properties of galaxies that is now accessible via large spectroscopic redshift surveys. In this chapter we focused on large filaments - as traced by galaxies and/or dark matter halos - in the local Universe. This is the regime that is now more readily accessible in observations and is also well supported by a solid understanding of structure formation on those scales, including the impact of large filaments on the growth of halos. Even so, key questions remain. For example, theory predicts that the cosmic web strongly modulates halo growth but it is still unclear what the impact of different halo assembly histories is on galaxies (at fixed halo or stellar mass), making it difficult to confront theory with observations. We also reviewed tension on the observed impact of large filaments on the neutral gas content of galaxies, making it difficult to link the large scale environment with gas reservoirs on a galaxy scale. This tension highlights difficulties in sample comparisons, but it also brings to fore the multi-scale aspect of the cosmic web and the potentially very different roles of different types of filaments. Being a second-order effect, the impact of the cosmic web on galaxies will always be difficult to isolate from stronger effects, such as halo or stellar mass and isotropic environment, making comparisons across different studies difficult and potentially creating tensions where there are none. There is also some unavoidable theoretical subjectivity and experimental uncertainty in the classification of a galaxy density field into distinct cosmic web environments. All of these difficulties, however, are also opportunities, forcing us to consider theoretical and observational uncertainties that can reveal what we think is a complicated but rich set of connections between different scales, galaxy types, and physical processes across all cosmic web environments, as well as redshifts. Forthcoming surveys will bring denser samples over wider redshift ranges, opening up plenty of opportunities for progress in developing a holistic view of galaxy formation and evolution within the context of large scale structure.

\begin{ack}[Acknowledgments]{}
The authors would like to thank the following PhD students for their time in reading the draft and providing comments and requests for clarifications, which improved the clarity of this article: Zoe Harvey, Baptiste Jego, Marta Ramos, Alfie Russell, Holly Seo, and Jinzhi Shen.
\end{ack}


\bibliographystyle{Harvard}
\bibliography{LSS-CW}

\end{document}